\documentclass[sigconf]{acmart}
\renewcommand\footnotetextcopyrightpermission[1]{} 
\usepackage{times}
\usepackage{soul}
\usepackage{url}
\usepackage[utf8]{inputenc}
\usepackage{graphicx}
\usepackage{amsmath}
\usepackage{amsthm}
\usepackage{booktabs}
\usepackage{algorithm}
\usepackage{algorithmic}
\usepackage[switch]{lineno}
\usepackage{makecell}
\usepackage{amsmath}
\usepackage{multirow}
\usepackage{arydshln}
\usepackage{subcaption}
\usepackage{threeparttable}
\usepackage{booktabs}
\usepackage{xurl}
\usepackage{hyperref}
\AtBeginDocument{%
  }

\setcopyright{acmlicensed}
\copyrightyear{2018}
\acmYear{2018}
\acmDOI{XXXXXXX.XXXXXXX}
\begin{document}

\title{Differential Mental Disorder Detection with Psychology-Inspired Multimodal Stimuli}

\author{Zhiyuan Zhou}
\email{zhouzhiyuan.hfut@gmail.com}
\affiliation{%
  \institution{Hefei University of Technology}
  \city{Hefei}
  \country{China}
}

\author{Jingjing Wu}
\email{2022800098@hfut.edu.cn}
\affiliation{%
  \institution{Hefei University of Technology}
  \city{Hefei}
  \country{China}
}

\author{Zhibo Lei}
\email{2023212130@mail.hfut.edu.cn}
\affiliation{%
  \institution{Hefei University of Technology}
  \city{Hefei}
  \country{China}
}

\author{Zhongcheng Yu}
\email{2024170763@mail.hfut.edu.cn}
\affiliation{%
  \institution{Hefei University of Technology}
  \city{Hefei}
  \country{China}
}

\author{Yuqi Chu}
\email{chuyuqi127@gmail.com}
\affiliation{%
  \institution{Hefei University of Technology}
  \city{Hefei}
  \state{Anhui Province}
  \country{China}
}

\author{Xiaowei Zhang}
\email{zhangxw@lzu.edu.cn}
\affiliation{%
  \institution{Lanzhou University}
  \city{Lanzhou}
  \state{Gansu Province}
  \country{China}
}

\author{Qiqi Zhao}
\email{zhaoqq21@lzu.edu.cn}
\affiliation{%
  \institution{Lanzhou University}
  \city{Lanzhou}
  \state{Gansu Province}
  \country{China}
}

\author{Qi Wang}
\email{2023010055@mail.hfut.edu.cn}
\affiliation{%
  \institution{Hefei University of Technology}
  \city{Hefei}
  \state{Anhui Province}
  \country{China}
}

\author{Shijie Hao}
\email{hfut.hsj@gmail.com}
\affiliation{%
  \institution{Hefei University of Technology}
  \city{Hefei}
  \state{Anhui Province}
  \country{China}
}

\author{Yanrong Guo}
\email{yrguo@hfut.edu.cn}
\affiliation{%
  \institution{Hefei University of Technology}
  \city{Hefei}
  \state{Anhui Province}
  \country{China}
}

\author{Richang Hong}
\email{hongrc.hfut@gmail.com}
\affiliation{%
  \institution{Hefei University of Technology}
  \city{Hefei}
  \state{Anhui Province}
  \country{China}
}

\renewcommand{\shortauthors}{Trovato et al.}

\begin{abstract}
Differential diagnosis of mental disorders remains a fundamental challenge in real-world clinical practice, where multiple conditions often exhibit overlapping symptoms. However, most existing public datasets are developed under single-disorder settings and rely on limited data elicitation paradigms, restricting their ability to capture disorder-specific patterns.
In this work, we investigate differential mental disorder detection through psychology-inspired multimodal stimuli, designed to elicit diverse emotional, cognitive, and behavioral responses grounded in findings from experimental psychology. Based on this paradigm, we collect a large-scale multimodal mental health dataset (MMH) covering depression, anxiety, and schizophrenia, with all diagnostic labels clinically verified by licensed psychiatrists.
To effectively model the heterogeneous signals induced by diverse elicitation tasks, we further propose a paradigm-aware multimodal framework that leverages inter-disorder differences prior knowledge as prompt-guided semantic descriptions  to capture task-specific affective and interaction contexts for multimodal representation learning in the new differential mental disorder detection task.
Extensive experiments show that our framework consistently outperforms existing baselines, underscoring the value of psychology-inspired stimulus design for differential mental disorder detection. \footnote{The full dataset and code will be publicly available soon.}
\end{abstract}




\maketitle

\section{Introduction}
Mental health disorders, such as depression (MD), anxiety (ANX), and schizophrenia (SC), are an escalating global public health concern, which affect hundreds of millions of individuals and impose a substantial burden on healthcare systems \cite{WHO2022MentalHealth}. Early detection and accurate diagnosis are critical for timely intervention and effective treatment. 
However, conventional clinical assessments largely rely on self-report questionnaires or face-to-face interviews, which are often time-consuming and subjective. 
Recent advances in deep learning have enabled multimodal mental health analysis using audio, video, and textual signals collected during structured elicitation tasks \cite{10.1145/3664647.3681491,4,11086398}. By modeling facial expressions, speech prosody, and linguistic content, these approaches offer a more objective and scalable alternative for supporting mental health diagnosis.

\begin{figure}[t]
    \centering
    \includegraphics[width=\linewidth]{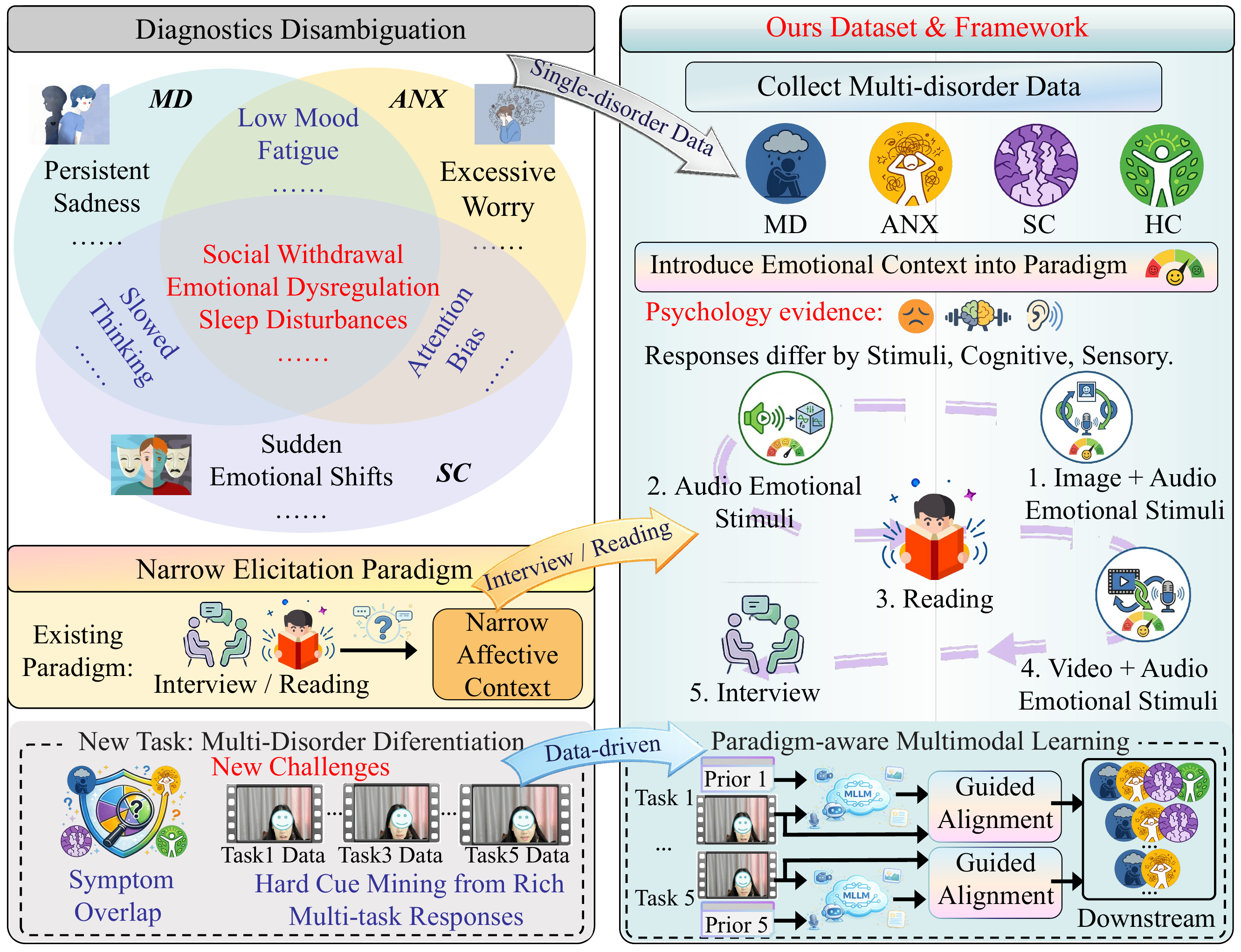} 
    \caption{Comparison between conventional single-disorder assessment based on interview- or reading-based paradigms and our multi-disorder differential diagnosis setting built on a psychology-inspired multimodal elicitation paradigm, together with the proposed paradigm-aware multimodal learning framework. MD, ANX, SC and HC refer to depression, anxiety,  schizophrenia and healthy controls, respectively.
  }
    \label{fig:intro}
\end{figure}

Despite these advances, existing multimodal mental health methods \cite{dibekliouglu2015multimodal,7763752,gui2019cooperative,an-etal-2020-multimodal,jung2024hique} remain limited in real-world diagnostic scenarios. Most current approaches \cite{10.1145/3664647.3681491,cheong2024fairrefuse} formulate mental disorder detection as a single-disorder task and improve performance primarily through data-driven modeling. Specifically, some methods \cite{fan2024transformer,chen2025text,10943944} focus on enhancing intra-modal feature modeling to capture discriminative modality-specific expressions, while others \cite{chen2024iifdd,zhou2024multi,shi2025hope} emphasize cross-modal feature fusion to combine complementary information across modalities.
They  are typically developed on datasets \cite{gratch2014distress,zou2022semi,pan2024disentangled} that focus on only one condition, most commonly depression. Such settings implicitly assume homogeneous diagnostic contexts and simplified decision boundaries.
In contrast, clinical practice requires differential diagnosis, as multiple mental disorders often coexist and exhibit overlapping emotional, cognitive, and behavioral symptoms, such as social withdrawal, emotional dysregulation, and impaired communication (Fig.~\ref{fig:intro}). This discrepancy between dataset construction, computational task design, and real-world diagnostic requirements substantially limits the practical applicability of existing methods.

\begin{table*}[h]
  \centering
  \caption{Comparison of our proposed dataset with existing public interview-based mental health datasets. The proposed dataset covers more disorder types: MD, ANX, and SC, and uses richer multi-modal triggering methods.}
    \begin{tabular}{llllll}
    \hline
    Dataset & Modal & Disorder Type & Sample Size & Triggering Method & Quantification Standards \\
    \hline\hline
    DAIC \cite{gratch2014distress}  & A, V, T & MD    & 189   & Interview & PHQ-8 \\
    E-DAIC \cite{devault2014simsensei} & A, V, T & MD    & 275   & Interview, Reading & PHQ-8 \\
    Pittsburgh \cite{dibekliouglu2017dynamic}& A, V, T & ANX   & 130   & Interview & HAMD + Clinical
Diagnosis \\
    MODMA \cite{modma}& A, V, EEG & MD    & 55    & Interview, Reading & PHQ-9 + Clinical
Diagnosis \\
    StudentSADD \cite{tlachac2022studentsadd}& A, T   & MD & 300   & Interview, Writing & PHQ-9 \\
    CMDC \cite{zou2022semi} & A, V, T & MD    & 78  & Interview & PHQ-9 + Clinical
Diagnosis\\
     \hline
    MMH (Ours)  & A, V, T & \textbf{MD/ANX/SC}   &\textbf{928}   & \textbf{\makecell{Listening, Watching, \\  Interview,  Reading }} & PHQ-9 + Clinical
Diagnosis  \\
    \hline
    \end{tabular}%
    
  \label{tab:addlabel}%
\end{table*}

A key but underexplored factor underlying the above limitation lies in how behavioral and emotional signals are elicited. Existing multimodal mental health datasets \cite{gratch2014distress,devault2014simsensei,modma} predominantly rely on interviews or reading-based tasks, which provide a narrow range of affective contexts and often fail to evoke disorder-specific symptom manifestations.
Experimental psychology studies \cite{fournier2013amygdala,ross2007impaired} suggest that different mental disorders respond divergently to variations in emotional stimuli, cognitive load, and sensory modalities. For example, depressive symptoms may manifest as blunted affect during passive observation, anxiety-related patterns may emerge under emotionally charged or evaluative situations, while schizophrenia-related impairments are often more evident in spontaneous communication and multimodal integration. These findings indicate that richer, psychology-grounded elicitation paradigms are critical for differential mental disorder detection.

Motivated by these disorder–context associations, we design a psychology-grounded multimodal elicitation paradigm  to collect data for differential mental disorder detection. The five elicitation tasks included in the proposed paradigm form a minimal yet sufficient set for emotion or behavior elicitation (Fig. \ref{fig:intro}): Multi-modal Stimulation I (image + audio) and Unimodal Stimulation (audio) probe affective induction, Text Reading provides a controlled speech baseline, Multi-modal Stimulation II (video + audio) captures higher-level social–emotional cognition, and the Human–Computer Interview elicits spontaneous verbal and interactive behaviors. This design ensures comprehensive coverage of emotional reactivity, cognitive load, social evaluation, and cross-modal integration, while remaining feasible for large-scale data collection.
Based on this paradigm, we construct a large-scale multimodal mental health dataset (MMH) comprising 928 participants (24,128 facial video clips and 14,848 audio–text pairs) and encompassing depression, anxiety, and schizophrenia. All diagnostic labels are clinically verified by licensed psychiatrists, ensuring high reliability for differential disorder detection.

Although multi-task psychology-inspired stimulation enhances the discriminability of disorder-specific symptoms, it also leads to more heterogeneous signals and makes discriminative cue mining from a large amount of collected data substantially more challenging. To handle the new task of differential mental disorder detection under diverse elicitation scenarios, we propose a paradigm-aware multimodal learning framework (PMLF). PMLF leverages disorder-specific prior knowledge to generate prompt-guided semantic descriptions for distinct stimulation tasks. These descriptions characterize task-specific affective and interaction contexts at the sample level. Based on them, the model further enhances multimodal feature alignment and fusion for multiple downstream tasks.
This design moves the model beyond the purely data-driven paradigm of prior single-disorder approaches \cite{wei2022multi,niu2020multimodal}. It allows the model to extract clinically meaningful evidence under explicit cross-disorder guidance from diverse tasks. This leads to better fine-grained differential detection and improved diagnostic interpretability.

In summary, our contributions are as follows:
\begin{itemize}
\item We introduce a psychology-inspired multimodal stimulus paradigm and construct a clinically verified multi-disorder multimodal dataset (MMH), enabling rich elicitation of disorder-specific affective and behavioral cues;
\item We propose a paradigm-aware multimodal learning framework (PMLF) that leverages prompt-guided semantic information of different elicitation tasks to enhance differential mental disorder detection;
\item Extensive experiments demonstrate the superiority of the proposed framework, and evaluate the value of psychology-grounded stimulus design for differential diagnosis of mental disorders.
\end{itemize}

\section{Related Works}

\label{mmddd}

\subsection{Multi-modal Mental Disorder Datasets} 

Several multimodal interview-based datasets have been introduced to support computational mental health assessment, which can be broadly organized by target disorder. 

For depression, representative datasets include DAIC and E-DAIC \cite{gratch2014distress,devault2014simsensei}, two widely used benchmarks that provide audio, video, and transcribed text collected from semi-structured interviews or additional reading tasks, with 189 and 275 subjects, respectively, and PHQ-8-based annotations. 
Other depression-related datasets include MODMA \cite{modma}, which provides audio, video, and EEG recordings from 55 participants with PHQ-9 scores. StudentSADD \cite{tlachac2022studentsadd} contains audio and text data from 300 students, collected through interview and writing tasks with depression- and suicidality-related annotations. CMDC \cite{zou2022semi} is a small-scale interview dataset with audio, video, and text modalities, annotated by PHQ-9 labels.
For anxiety, the Pittsburgh dataset \cite{dibekliouglu2017dynamic} is a representative benchmark, containing audio and video data from 130 subjects collected through structured interviews and annotated with the Hamilton Anxiety Rating Scale. 
For schizophrenia, multimodal behavioral data are particularly valuable due to the difficulty of data collection and the limited accessibility of patient resources. To the best of our knowledge, no publicly available multimodal benchmark currently exists for schizophrenia assessment. 

While these datasets have significantly advanced the field, they have two common limitations: (1) most focus on a single disorder, and (2) behavioral signals are predominantly elicited through homogeneous paradigms, such as interviews or reading-based tasks, which restrict the diversity of affective and cognitive contexts.
To address these limitations, our proposed dataset incorporates multiple mental health disorders and is structured around diverse psychology-inspired elicitation scenarios. A detailed comparison with existing public datasets is provided in Table \ref{tab:addlabel}.

\subsection{Multi-modal Mental Disorder Detection Methods} 

Multi-modal mental health detection systems integrate heterogeneous signals, including audio/speech, visual or facial behavior, text/language, and physiological responses, to capture complementary aspects of mental health conditions. 
Specifically, for the visual modality, many studies \cite{gong2017topic,chen2024depression,flores2022temporal} extract facial behavior cues using tools such as OpenFace \cite{openface}, including facial action units, gaze, and head-motion features. 
For the audio modality, commonly used representations include handcrafted acoustic descriptors such as Mel-Frequency Cepstral Coefficients (MFCCs) \cite{davis1980comparison} as well as pretrained speech features like wav2vec~2.0 \cite{baevski2020wav2vec}. 
For the text modality, most methods \cite{rodrigues2019multimodal,fan2019multi,10943944} employ BERT-based encoders \cite{devlin2019bert} to obtain contextual semantic representations. 
By jointly modeling multiple modalities, these systems have demonstrated clear advantages over single-modality approaches in providing more comprehensive evidence for mental health assessment.

Existing multimodal methods for depression detection can be broadly grouped into two categories.  
One focuses on intra-modal feature learning, which aims to capture discriminative temporal and semantic patterns within each modality \cite{fan2019multi,fan2024transformer,10943944}. 
Based on these modality-specific representations, some studies further focus on intra-modal feature enhancement.
For instance, Uddin et al. \cite{uddin2022deep} develop a spatio-temporal framework to model facial and linguistic cues for depression assessment. 
The other emphasizes cross-modal feature integration, with the goal of effectively exploiting complementary information across different modalities \cite{morales2018linguistically,chen2024iifdd,shi2025hope}. 
More recent methods \cite{fan2024transformer,chen2024iifdd,mou2025disentangled} further adopt attention-based or Transformer-based fusion modules to explicitly model cross-modal interactions and exploit complementary information across modalities.
A representative example is Niu et al. \cite{niu2020multimodal}, who propose a spatio-temporal attention network coupled with a multimodal attention-based fusion strategy to learn joint representations for depression prediction.

Similarly, anxiety-related multimodal methods can also be broadly grouped into these two categories, i.e., intra-modal feature learning and cross-modal feature integration.
For the former, Diep et al. \cite{diep2022multi} collect speech and text data from self-management tasks and combine both deep and handcrafted features for anxiety-related assessment. 
For the latter, a representative example is Tsai et al. \cite{tsai2019multimodal}, who introduce a multimodal Transformer with cross-modal attention to capture temporal dependencies and cross-modal interactions for mental health analysis.

Compared with depression and anxiety, multimodal research on schizophrenia remains much more limited. 
Existing studies \cite{wang2022adaptive,li2023diagnosis} have traditionally focused on neuroimaging data, such as MRI, while behavior-based approaches \cite{fan2019multi,ray2019multi} using audio, visual, and textual modalities have only recently begun to emerge. 
For instance, work in \cite{6} develops an attention-based fusion framework over audio, video, and text for schizophrenia analysis. 
However, this non-public dataset contains only 7 schizophrenia samples, reflecting both the rarity of such cases and the difficulty of collecting publicly accessible data.

Despite these advances, most existing methods are developed for single-disorder binary classification. The shift to multi-disorder classification and richer multi-task data introduces new challenges, making it harder for existing methods to mine discriminative cues among disorders.
Therefore, we propose a paradigm-aware multimodal learning framework to better mine disorder-discriminative cues from rich multi-task data.

\begin{figure*}[h]
    \centering
    \includegraphics[width=\textwidth]{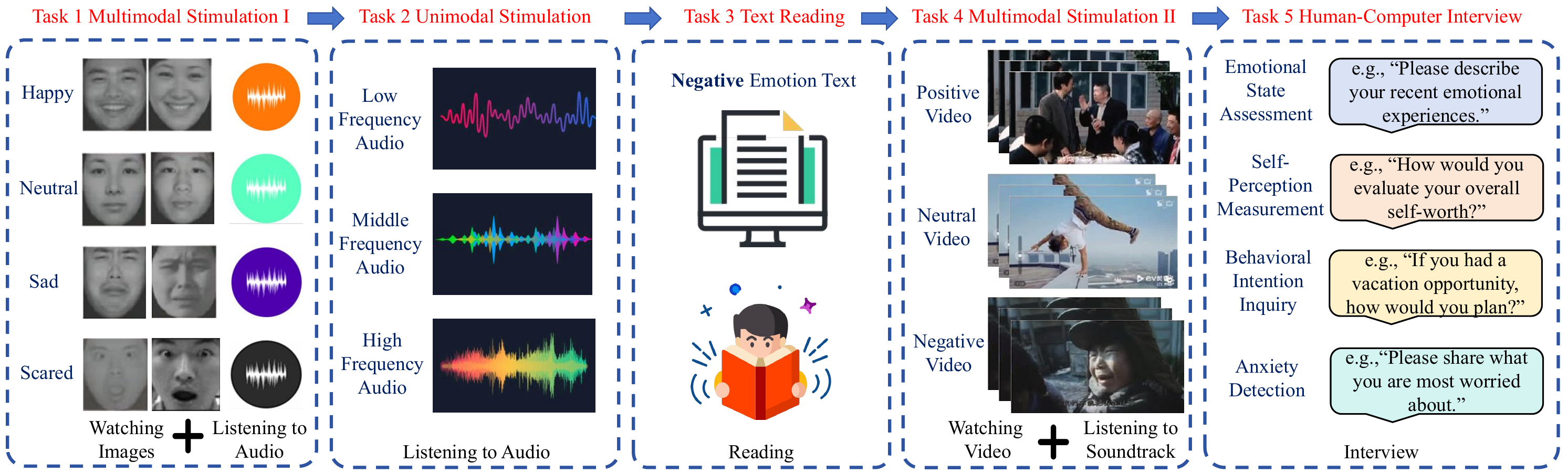}
    \caption{
    Overview of the psychology-inspired multimodal stimulus paradigm.
    }
    \label{fig:data_paradigm}
\end{figure*}

\section{Psychology-Inspired Multimodal Mental Health Dataset (MMH)}
To support differential mental disorder detection under diverse affective and cognitive contexts, we construct a large-scale multimodal mental health (MMH) dataset based on a psychology-inspired elicitation paradigm, covering depression (MD), anxiety (ANX), and schizophrenia (SC).

\subsection{Psychology-Inspired Multimodal Elicitation Paradigm}
We design a psychology-inspired multimodal elicitation paradigm to capture disorder-specific behavioral and emotional patterns under heterogeneous affective and cognitive contexts. The paradigm records synchronized facial expression video and audio signals from participants across multiple triggering tasks using a non-contact and rapid acquisition setup, enabling scalable data collection with minimal participant burden. Based on findings from experimental psychology, the protocol integrates both multimodal and unimodal emotional stimuli to elicit complementary affective and communicative cues across different interaction settings. Each participant follows a fixed five-step elicitation procedure to ensure experimental consistency and control. As illustrated in Fig.~\ref{fig:data_paradigm}, the  five structured elicitation tasks are described as follows:

\textbf{Part 1: Multi-modal Stimulation I (MS-I).}
This module aims to assess participants’ emotional reactivity, attention allocation, and multisensory integration under socially relevant affective contexts. Following findings from experimental psychology \cite{schirmer2006beyond}, MS-I presents synchronized visual and auditory stimuli to simulate real-life social perception. The module consists of four clips corresponding to four emotional categories: happy, neutral, sad, and scared. Each clip includes two images and one audio segment. The images are selected from the International Affective Picture System (IAPS) and are displayed for 15 seconds.

\textbf{Part 2: Unimodal Stimulation (US).}
This module focuses on isolating auditory emotional processing, including speech prosody perception and affective semantic understanding. Based on established auditory emotion perception studies \cite{banse1996acoustic}, US presents audio-only emotional stimuli. The module contains nine audio clips grouped into three frequency bands (low, middle, and high). Each clip lasts 15 seconds, with continuous presentation within each group and a 5-second pause between groups.

\textbf{Part 3: Text Reading.}
This task aims to evaluate basic speech production abilities and acoustic characteristics under controlled semantic conditions. Participants are instructed to read a paragraph with negative emotional content. The task is designed to capture acoustic features such as pitch, speech rate, and articulation patterns, enabling assessment of vocal expressiveness and fluency.

\textbf{Part 4: Multi-modal Stimulation II (MS-II).}
This module evaluates higher-level affective and social cognition, including emotional recognition, social interpretation, regulatory capacity, and multimodal affective integration. Following experimental psychology paradigms on facial emotion processing \cite{gur1992facial}, MS-II presents emotional video clips in a fixed order: positive, neutral, and negative. Each clip is followed by a 5-second break to allow emotional transition and stabilization.

\textbf{Part 5: Human–Computer Interview.}
This task assesses verbal expression, emotional reactivity, and social interaction patterns under structured conversational settings. Similar to existing mental health datasets \cite{gratch2014distress,devault2014simsensei,modma}, a structured interview is conducted via a human–computer dialogue system. The session consists of 15 standardized questions designed based on clinical diagnostic criteria (e.g., DSM-V), covering four dimensions: (1) Emotional State Assessment, (2) Self-Perception Measurement, (3) Behavioral Intention Inquiry, and (4) Anxiety Detection.

\begin{table}[t]
    \centering
    \caption{Statistics of the dataset.}
    \renewcommand{\arraystretch}{1.05}
    \setlength{\tabcolsep}{6pt}
    \begin{tabular}{llr}
        \toprule
        & \textbf{Item} & \textbf{Stats} \\
        \midrule

        \multirow{5}{*}{\textbf{Sample}}
            & Train Set & 557 \\
            & Val Set & 186 \\
            & Test Set  & 185 \\
            & Total     & 928 \\
        \cdashline{2-3}
            & Video Segments Per Sample   & 26 \\
            & Audio/Text Segments Per Sample    & 16 \\
        \cdashline{1-3}

        \multirow{5}{*}{\textbf{Modality}}
            & Video Segments & 24128 \\
            & Audio/Text Segments & 14848 \\
            
        \cdashline{2-3}
            & Avg. Length (Sec) Per Aud/Vid & 23.84 \\
            & Avg. Words Per Sample & 396.72 \\
        \cdashline{1-3}

        \multirow{4}{*}{\textbf{Labels}}
            & MD (Male/Female)  & 26/41 \\
            & SC (Male/Female)  & 38/11 \\
            & ANX (Male/Female)  & 26/86 \\
            & HC (Male/Female)  & 514/186 \\
        \bottomrule
    \end{tabular}
    \label{tab:mgd_stats}
    \vspace{-4mm}
\end{table}

\begin{figure*}[h]
    \centering
    \includegraphics[width=0.98\textwidth]{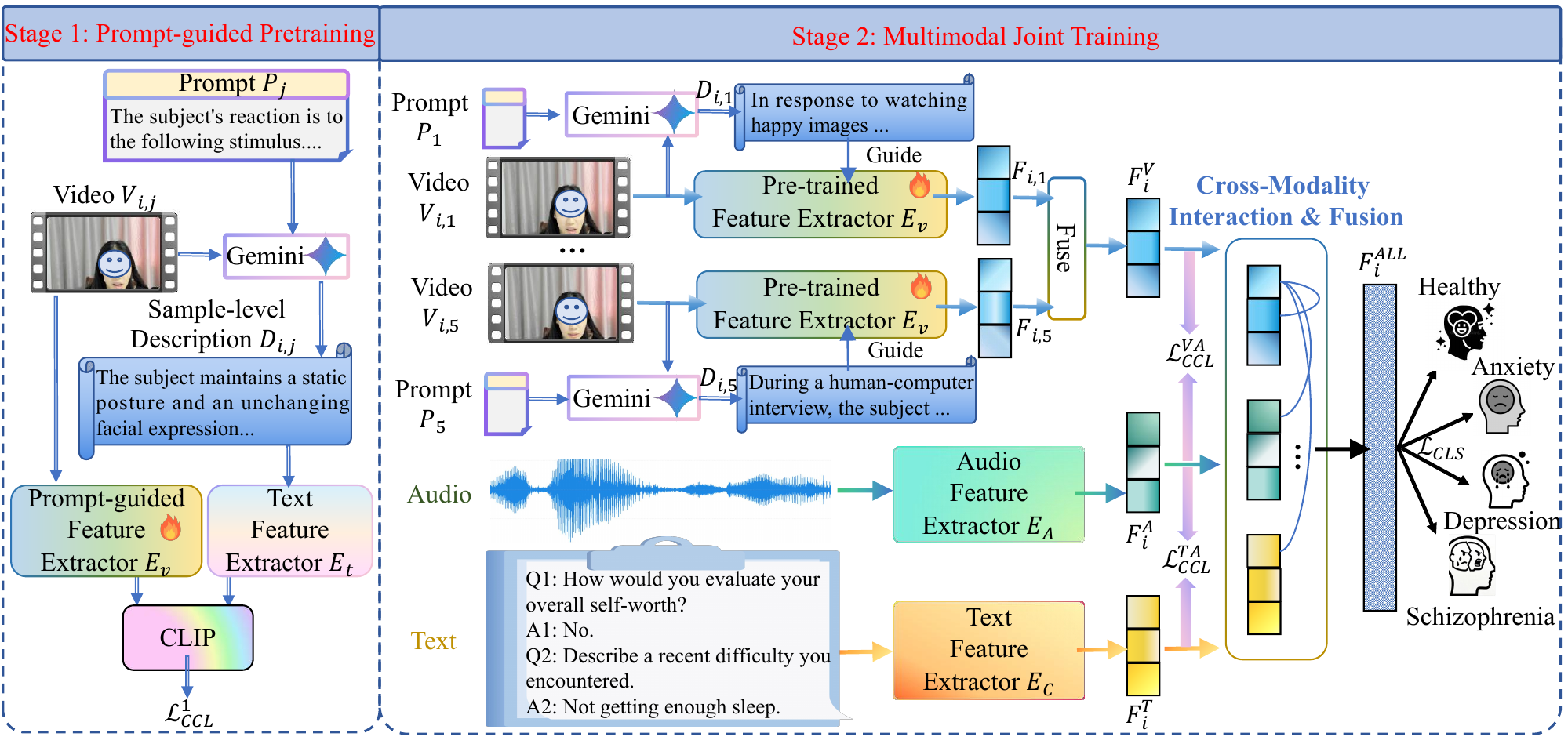} 
    \caption{
    The overview of the proposed paradigm-level prompt-guided learning framework, taking the four-class downstream task as an example.
    Stage 1 pretrains a video feature extractor by aligning visual cues with paradigm-level semantic descriptions generated by a MLLM. Stage 2 integrates these paradigm-aware visual features with complementary audio and text modalities through a cross-modality interaction module for multi-disorder detection.}
   \label{fig:overview}
\end{figure*}

\subsection{Dataset Statistics}
Based on the proposed psychology-inspired multimodal elicitation paradigm, we construct a large-scale multimodal mental health (MMH) dataset consisting of 928 participants and three modalities: video, audio, and text. For each participant, facial expression videos are recorded throughout the entire acquisition process, resulting in 26 video clips, together with 16 audio recordings collected during the text reading and human–computer interview stages; the text modality is obtained via transcription of the corresponding audio. The dataset includes four groups: 700 healthy individuals, 67 with depression (MD), 112 with anxiety (ANX), and 49 with schizophrenia (SC). The class distribution reflects the prevalence
characteristics and practical recruitment difficulty of different
populations in real-world mental health screening scenarios, especially
for clinically diagnosed patient groups. All patients diagnosed with MD, ANX, and SC are clinically validated by licensed psychologists to ensure label reliability. Participant ages range from 10 to 59 years, and the gender distribution of each group is summarized in Table~\ref{tab:mgd_stats}. Note that all participants have signed informed consent forms.


\section{Methodology}

\subsection{Overview}
Motivated by the observation that different elicitation paradigms induce distinct yet subtle behavioral and emotional patterns, we propose a paradigm-aware multimodal learning framework (PMLF) for differential mental disorder detection. Unlike conventional approaches that rely primarily on low-level facial geometry, our framework is designed to capture paradigm-level semantic information induced by diverse interaction settings. As illustrated in Fig.~\ref{fig:overview}, the framework consists of two stages: (1) \textit{Prompt-guided  pretraining}, where we leverage multimodal large language model (MLLM), specifically Gemini-2.5-pro \cite{gemini25flash}, to generate paradigm-aware textual descriptions that serve as semantic anchors for pretraining a video feature extractor via cross-modality contrastive learning; and (2) \textit{Multimodal joint training}, where the pretrained video representations are integrated with audio and text modalities through a modality interaction and fusion module to aggregate complementary information across paradigms for final multi-disorder classification.

\subsection{Stage 1: Prompt-guided Pretraining}

Learning discriminative video representations for mental disorder detection is challenging, because behavioral cues in facial videos are often subtle, sparse, and highly dependent on the elicitation paradigm. The same individual may exhibit different affective and interaction patterns across tasks, while visually similar facial behaviors may correspond to different underlying psychological states. As a result, relying solely on low-level facial movements or purely data-driven visual modeling may fail to capture clinically meaningful task-specific semantics.
To address this issue, we introduce paradigm-aware semantic supervision for video representation learning. The key idea is to guide the video encoder not only with facial behavior signals, but also with high-level textual descriptions that explicitly characterize the emotional, cognitive, and interaction context induced by each elicitation task.

Given a sample $i$, its video $V_{i,j}$ is collected under elicitation task $j$ ($j \in \{1,2,\dots,5\}$). We first employ OpenFace~\cite{openface} to extract frame-level facial behavior features following common processing practices in existing datasets and methods, including facial landmarks, head pose, gaze direction, and facial action units. These low-level behavioral signals are then fed into a backbone network to learn higher-level and paradigm-aware video representations, which are encoded by a prompt-guided video feature extractor $E_v$.

In parallel, for each video $V_{i,j}$, we leverage Gemini-2.5-pro~\cite{gemini25flash} to generate a \textit{sample-level description} $D_{i,j}$ based on a task-specific prompt $P_j$. The generated description captures high-level emotional, cognitive, and interaction semantics induced by the corresponding elicitation task, and is encoded by a text encoder $E_t$.

To bridge low-level visual signals and high-level semantic descriptions, we adopt CLIP~\cite{clip} to model cross-modal correlations between video and text modalities, and apply Cross-Modality Contrastive Learning (CCL)~\cite{ccl} to align the embeddings $z_{i,j}^v = E_v(V_{i,j})$ and $z_{i,j}^d = E_t(D_{i,j})$ in a shared latent space:

\begin{equation}
\mathcal{L}_{CCL}^{1} = - \log \frac{\exp(\mathrm{sim}(z_{i,j}^v, z_{i,j}^d)/\tau)}{\sum_{\substack{(m,n)\neq(i,j)}} \exp(\mathrm{sim}(z_{i,j}^v, z_{m,n}^d)/\tau)}
\end{equation}

where $\mathrm{sim}(\cdot)$ denotes cosine similarity, $z_{m,n}^d$ denotes the textual feature for sample $m$ under task $n$, and $\tau$ is a temperature parameter set to 0.2 following CCL~\cite{ccl}. This alignment encourages the video encoder to capture fine-grained emotional cues and interaction contexts that are explicitly described in the paradigm-aware textual supervision.

\begin{table*}[h]
\centering
\caption{Comparison of methods across four categories (HC/MD/SC/ANX) and overall performance. The best results are highlighted in \textbf{bold}, and the second-best results are underlined. 'w/' denotes 'with'.}
\label{tab:multiclass}
\setlength{\tabcolsep}{4pt}
\renewcommand{\arraystretch}{1}
\begin{threeparttable}
\resizebox{\textwidth}{!}{%
\begin{tabular}{l | c |  c c c c c c c}
\hline
\cline{3-9}
 Method & Modality & ACC & Macro-P & Macro-R & Macro-F1 & Weighted-P & Weighted-R & Weighted-F1 \\
\hline

GPT-4o~\cite{gpt4o} & T  & 17.20 & 38.54 & 45.18 & 23.79 & 72.55 & 17.21 & 15.59 \\
Grok-4\footnotemark[2]  & T  & 14.52 & 25.55 & 37.98 & 14.77 & 53.25 & 14.51 & 13.32 \\
Gemini-3~\cite{team2023gemini} & T  & 57.53 & 33.17 & 49.82 & 35.56 & 66.23 & 57.53 & 59.70 \\
\hdashline

GPT-4o~\cite{gpt4o} & VT  & 10.75 & 21.14 & 23.81 & 7.13 & 54.09 & 10.75 & 8.71 \\
Gemini-2.5-flash~\cite{gemini25flash} & VT  & 32.26 & 30.85 & 42.92 & 23.09 & 72.65 & 32.26 & 38.51 \\
Grok-4\footnotemark[2] & VT  & 11.29 & 27.59 & 25.66 & 10.78 & 61.62 & 11.29 & 9.95 \\
Doubao-1.5-vision-pro ~\footnotemark[3] & VT  &25.81 & 51.6 & 34.11 & 22.06 & 79.14 & 25.81 & 31.62 \\
Qwen2.5-vl-7B-instruct ~\cite{bai2025qwen25vltechnicalreport} & VT  & 20.43 & 54.26 & 32.53 & 19.78 & 78.98 & 20.43 & 22.85 \\
Qwen2.5-vl-32B-instruct ~\cite{bai2025qwen25vltechnicalreport} & VT & 10.22 & 22.84 & 25.89 & 5.44 & 63.33 & 10.21 & 6.28 \\
\hdashline
Zhou et al.~\cite{11086398}& AT  & 88.17 & 62.94 & 56.36 & 53.49 & 85.43 & 88.17 & 85.10 \\
Fan et al.~\cite{fan2024transformer} & AT & 87.10 & 36.96 & 50.00 & 41.18 & 80.93 & 87.10 & 82.92 \\
Depmamba~\cite{ye2025depmamba} & AT  &88.17 & 50.66 & 57.73 & 52.78 & 84.18 & 88.17 & 85.54 \\
\hdashline
Fan et al.~\cite{fan2024transformer} & AVT  & 88.17 & 51.10 & 60.71 & 54.46 & 85.38 & 88.17 & 86.34 \\
Depmamba~\cite{ye2025depmamba} & AVT  & 87.63 & 62.22 & 52.5 & 45.96 & 86.43 & 87.63 & 84.01 \\
Kumar et al. ~\cite{10943944} & AVT  & 88.71 & 61.13 & 58.8 & 58.03 & 87.63 & 88.71 & 87.60 \\
DNet~\cite{jiang2025dnet} & AVT  & 89.25 & 59.10 & 60.00 & 56.37 & 86.24 & 89.25 & 86.67 \\
\hdashline
PMLF w/ Transformer & AVT  & 89.78 & 54.07 & 62.18 & 57.63 & 86.03 & 89.78 & 88.02 \\
PMLF w/ Mamba & AVT & 90.32 & \underline{74.51} & 66.23 & 67.42 & \underline{90.86} & 90.32 & 89.67 \\
PMLF w/ Resnet & AVT  & \underline{91.40} & 70.85 & \underline{69.94} & \underline{68.13} & 90.22 & \underline{91.40} & \underline{90.01} \\
PMLF w/ (Resnet + Transformer) & AVT & \textbf{91.94} & \textbf{77.58} & \textbf{72.44} & \textbf{69.97} & \textbf{92.12} & \textbf{91.94} & \textbf{90.37} \\
\hline
\end{tabular}%
}
\end{threeparttable}
\end{table*}

\subsection{Stage 2: Multimodal Joint Training}
In Stage 2, we aim to fully exploit the diverse paradigms and modalities by jointly modeling their interactions.
For each elicitation task $j$, a dedicated video feature extractor, initialized from Stage 1, is employed to extract task-specific visual features $F_{i,j}$. Each extractor is guided by its corresponding paradigm-aware textual description to maintain semantic consistency, obtaining $F_i^V$. By leveraging these descriptions, the model can more effectively discriminate task-specific behavioral cues.
Following the processing pipeline used in existing datasets and methods, for audio modality, we extract 13-dimensional Mel-Frequency Cepstral Coefficients (MFCCs) \cite{davis1980comparison} to characterize spectral properties. These are processed by an audio backbone to yield the representation $F_i^A$.
For text modality, transcribed speech from interviews is encoded using BERT-base-Chinese \cite{devlin2019bert} to extract contextualized embeddings, which are further refined by a text backbone to yield $F_i^T$.


To maintain semantic consistency across modalities, we further apply cross-modality contrastive learning between $F_i^{V}$ and  $F_i^{A}$ with $\mathcal{L}_{CCL}^{VA}$ and $F_i^{T}$ and  $F_i^{A}$ with $\mathcal{L}_{CCL}^{TA}$.  $\mathcal{L}_{CCL}^{VA}$  and $\mathcal{L}_{CCL}^{TA}$ are computed with Eq. 1. The aligned features $\{F_i^V, F_i^A, F_i^T\}$ are then fed into a Modality Interaction and Fusion Module, which utilizes cross-attention mechanisms \cite{trans} to explicitly model inter-modality dependencies, integrating task-specific cues into a unified representation $F_i^{ALL}$ for multi-class mental health classification.
For mental health state prediction, we employ a standard cross-entropy loss on the fused feature $F_i^{ALL}$.
\begin{equation}
\mathcal{L}_{CLS} = - \sum_{c=1}^{C} y_c \log \hat{y}_c
\end{equation}
where $C$ is the number of mental health categories, $y_c$ is the ground-truth label, and $\hat{y}_c$ denotes the predicted probability.
The total loss for Stage 2 is defined as the equal sum of $\mathcal{L}_{CLS}$, $\mathcal{L}_{CCL}^{VA}$ and $\mathcal{L}_{CCL}^{TA}$.

\section{Experiments}
\subsection{Implementation Details}
The proposed models are implemented using PyTorch 1.12 and trained on an Nvidia GTX A6000 GPU. 
We maintain the original class distribution in the training, validation, and test splits to ensure a fair evaluation.
All models are trained for 80 epochs using the AdamW optimizer \cite{adam} with a learning rate of $1\times10^{-3}$ and a batch size of 12. Performance is evaluated using Accuracy (ACC), Precision (P), Recall (R), and F1-score (F1), reported in both macro and weighted forms.

\footnotetext[2]{\url{https://x.ai/news/grok-4}}
\footnotetext[3]{\url{https://console.volcengine.com/}}

\subsection{Differential Diagnosis of Mental Disorders}

In Table~\ref{tab:multiclass}, we report the results of our PMLF with various backbones, including ResNet \cite{10943944}, Transformers \cite{fan2024transformer}, hybrid ResNet-Transformer \cite{jiang2025dnet} and the Mamba architecture \cite{ye2025depmamba}. We compare the proposed PMLF with some reproduced mental disorder detection methods, as well as with state-of-the-art general LLMs and MLLMs. Several important observations can be drawn.

First, LLMs exhibit limited capability for differential mental disorder detection. 
Although models such as GPT-4o, Grok-4, and Gemini-3 demonstrate strong general reasoning abilities, their performance remains consistently low across all disorder categories. This indicates that mental disorder detection is inherently challenging when relying solely on textual descriptions, as many clinically relevant cues are expressed through subtle vocal, facial, and behavioral patterns that cannot be fully captured by language alone.

Second, while multimodal perception is more consistent with clinical diagnostic practice, current MLLMs still struggle to effectively integrate long-duration video and text signals for multi-disorder detection. Models such as GPT-4o, Gemini, Grok, Doubao and Qwen exhibit unstable performance across different disorder categories. In some cases, introducing visual modality even leads to degraded performance compared to text-only settings, suggesting that existing MLLMs are not yet capable of reliably extracting and aggregating clinically meaningful cues from complex, long-range multimodal sequences.

Third, supervised task-specific models \cite{fan2024transformer,ye2025depmamba,10943944,jiang2025dnet,11086398} trained on our dataset exhibit more stable and reliable performance than general-purpose LLMs and MLLMs. Architectures such as  DNet all achieve competitive results, without the severe performance fluctuations observed in LLMs/MLLMs. These results indicate that task-specific training on the proposed dataset is essential for learning discriminative representations for differential mental disorder detection.

Finally, our method achieves the best overall performance. Across different backbone settings, PMLF consistently outperforms all compared methods, demonstrating both its general effectiveness and its compatibility with diverse architectures. These results verify the benefit of paradigm-aware semantic supervision and cross-modal representation alignment, and highlight the importance of jointly modeling elicitation contexts and multimodal interactions for differential mental disorder detection. In addition, weighted metrics are generally higher than macro metrics, mainly due to class imbalance in the dataset. This suggests that existing methods remain more effective on majority classes, while performance on minority classes still needs further improvement.

\begin{table}[t]
\caption{Ablation study on different elicitation paradigms.}
\label{tab:paradigm_ablation}
\centering
\resizebox{\columnwidth}{!}{
\begin{tabular}{l c c c c c}
\hline
Paradigm & ACC & Macro-P & Macro-R  & Macro-F1 & Weighted-F1\\
\hline
w/o MS-I        & 90.32 & 63.05 & 68.38 & 64.44 & 88.78 \\
w/o US          & 90.32	& 60.53	& 65.00	& 60.68	& 87.60  \\
w/o Reading     & 87.63	& 62.22	& 52.50	& 45.96	& 84.01 \\
w/o MS-II       & 89.78 & 56.67 & 63.86 & 59.21 & 87.38 \\
w/o Interview   & 90.32 & 61.29 & 65.00 & 61.29 & 87.88 \\
\hdashline
Ours            & 91.94 & 77.58 & 72.44 & 69.97 & 90.37 \\
\hline
\end{tabular}
}
\end{table}

\subsection{Key Research Questions}

\subsubsection{Effectiveness of Different Elicitation Tasks}
To analyze the contribution of different elicitation tasks, we conduct an ablation study by removing each task from the proposed paradigm. The results are reported in Table~\ref{tab:paradigm_ablation}.

Overall, removing any elicitation task degrades performance, confirming that different paradigms provide complementary information for differential mental disorder detection. In terms of Macro-F1 metric, removing the text reading task causes the largest drop, indicating that it contributes particularly important discriminative cues, likely because it requires subjects to process a relatively long passage with negative emotional content. Removing US, MS-II, or the interview tasks also leads to substantial degradation, while removing the MS-I results in moderate but consistent declines.
These results suggest that incorporating multiple and diverse elicitation tasks is beneficial for improving multi-disorder detection performance.

\subsubsection{Flexibility of MMH Across Different Diagnosis Tasks}
\label{fadd}

We further investigate the flexibility of MMH across different diagnosis settings. As shown in Table \ref{tab:table5}, MMH supports unified evaluation under both multi-class and pairwise classification scenarios, demonstrating its versatility for diverse clinical diagnosis needs. The varying performance across different disorder combinations suggests that MMH captures heterogeneous levels of inter-disorder similarity and task difficulty, making it a practical benchmark for studying diagnosis under different granularities. Here, the baseline adopts the same feature extraction backbone as our method, i.e., a ResNet-Transformer architecture, and performs direct classification without paradigm-aware modeling. Meanwhile, the proposed PMLF consistently improves over this baseline in almost all settings, with particularly large gains on the MDD-vs-SC and ANX-vs-SC tasks.
These results not only verify the effectiveness of PMLF in exploiting paradigm-aware discriminative cues, but also further confirm that MMH provides sufficiently rich and flexible data support for both general multi-disorder diagnosis and customized disorder-specific evaluation.

\begin{table}[t]
\centering
\caption{Results of different methods on three-class and binary differential diagnosis (\%).}
\label{tab:table5}
\resizebox{\columnwidth}{!}{%
\begin{tabular}{@{}l|ccc|ccccc@{}}
\hline
Method & MDD & ANX & SC & ACC & Macro-P & Macro-R & Macro-F1 & Weighted-F1\\
\hline
\multirow{4}{*}{Baseline} 
& $\checkmark$ & $\checkmark$ & $\checkmark$ & 63.04 & 59.17 & 56.41 & 56.92 & 61.37 \\
& $\checkmark$ & $\checkmark$ &              & 69.44 & 70.20 & 63.31 & 63.03 & 66.45 \\
& $\checkmark$ &              & $\checkmark$ & 66.67 & 70.00 & 61.43 & 59.66 & 62.46 \\
&              & $\checkmark$ & $\checkmark$ & 78.12 & 75.71 & 70.45 & 71.96 & 76.89 \\
\hline
\multirow{4}{*}{\makecell{Ours}} 
& $\checkmark$ & $\checkmark$ & $\checkmark$ & 65.22 & 64.44 & 61.56 & 62.21 & 64.08 \\
& $\checkmark$ & $\checkmark$ &              & 69.44 & 68.52 & 64.61 & 64.86 & 67.68 \\
& $\checkmark$ &              & $\checkmark$ & 83.33 & 84.38 & 81.43 & 82.22 & 82.96 \\
&              & $\checkmark$ & $\checkmark$ & 84.38 & 84.86 & 77.73 & 79.97 & 83.49 \\
\hline
\end{tabular}%
}
\end{table}

\subsubsection{Contribution of Different Modalities}

\begin{figure}[t]
    \centering
    \includegraphics[width=0.95\linewidth]{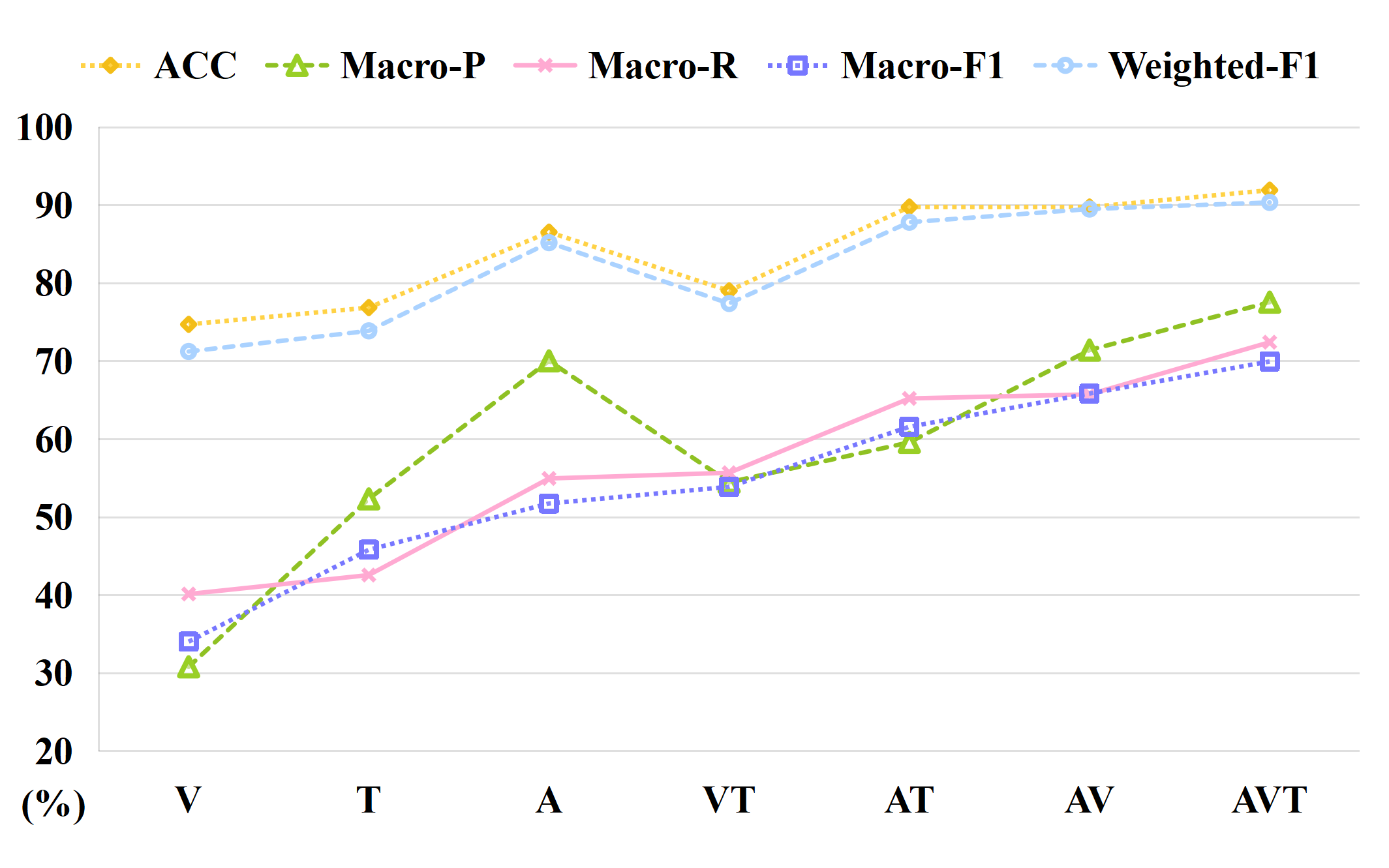} 
    \caption{Modality-level ablation study. 
  }
    \label{tab:modality_ablation}
\end{figure}

To evaluate the contribution of different modalities, we perform a modality-level ablation study with different modality combinations, as shown in Fig.~\ref{tab:modality_ablation}. Among single-modality settings, audio achieves the best performance, highlighting the importance of vocal cues for mental disorder detection, while video performs substantially worse, suggesting that facial expressions alone are insufficient under complex elicitation contexts. Multi-modality combinations consistently outperform single-modality settings. In particular, A+V significantly improves over audio-only and video-only models, demonstrating the complementarity between acoustic and facial cues. The full multimodal setting (A+V+T) achieves the best overall performance, confirming the benefit of jointly modeling facial, vocal, and linguistic signals. These results verify the effectiveness of multimodal fusion for robust multi-disorder detection.

\subsubsection{Module-Level Ablation}

\begin{figure}[t]
    \centering
    \includegraphics[width=0.9\linewidth]{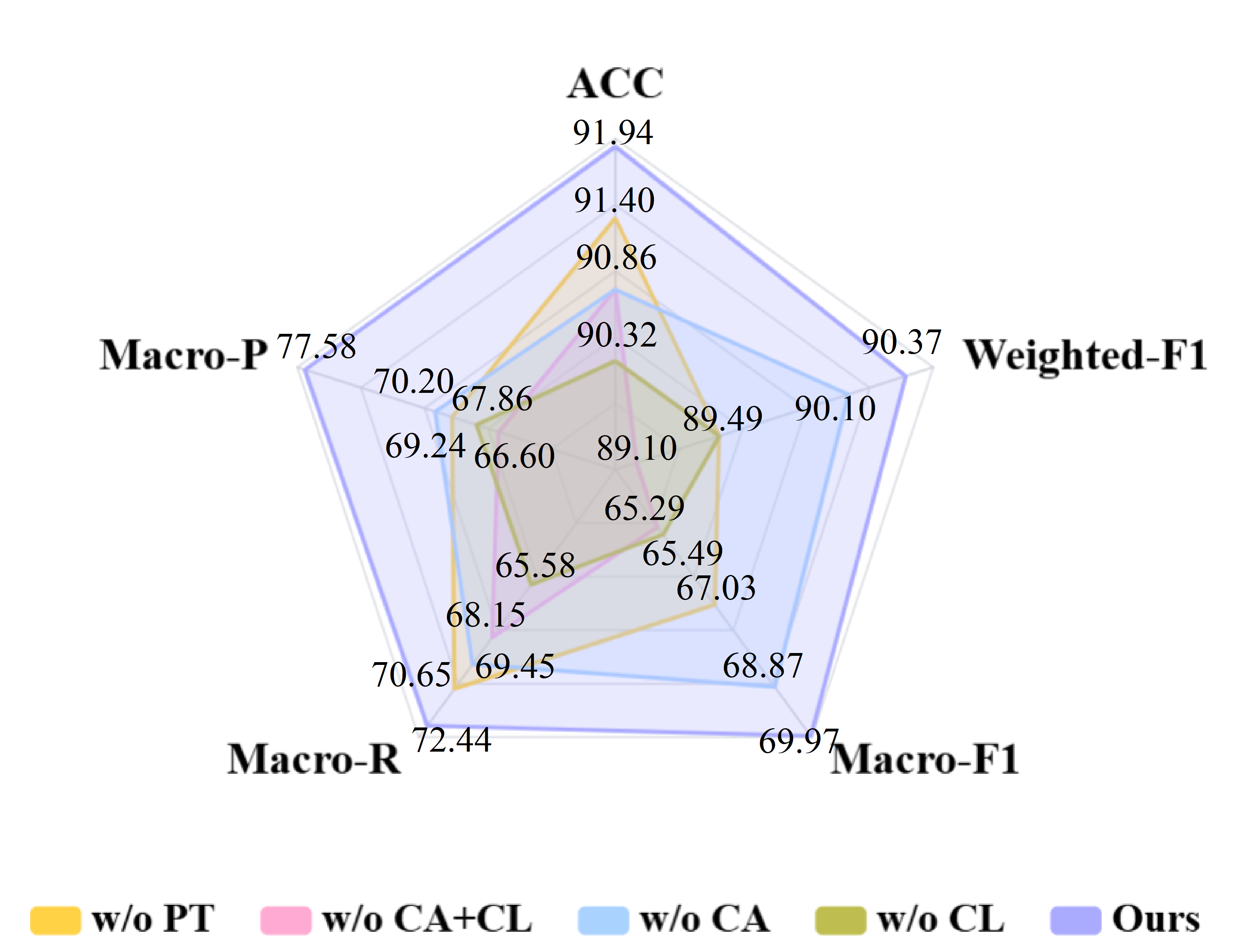} 
    \caption{Module-level ablation study on the paradigm-aware multimodal learning framework. (PT: Pretraining; CA: Cross-Attention; CL: Contrastive Learning) 'w/o' denotes 'without'.
  }
    \label{tab:module_ablation}
\end{figure}

To evaluate the contribution of each component in the proposed paradigm-aware multimodal learning framework, we conduct a module-level ablation study by removing individual modules and comparing the resulting performance with the full model. The results are summarized in Fig. ~\ref{tab:module_ablation}.

The full model (Ours) achieves the best overall performance, demonstrating the effectiveness of jointly optimizing all components. Removing the prompt-guided pretraining stage (w/o PT) leads to a noticeable performance drop, indicating that paradigm-aware semantic supervision is beneficial for initializing the video feature extractor with high-level contextual representations.

When both cross-attention and contrastive learning are removed (w/o CA+CL), the F1 score further decreases, highlighting the importance of cross-modal interaction and representation alignment for learning discriminative multimodal features. Among the individual components, removing cross-attention (w/o CA) results in a larger degradation than removing contrastive learning alone (w/o CL), suggesting that explicit cross-modal interaction plays a more critical role in integrating heterogeneous signals induced by different elicitation paradigms.

\begin{figure}[t]
\centering

\begin{minipage}[t]{0.45\textwidth}
    \centering
    \includegraphics[width=0.45\linewidth]{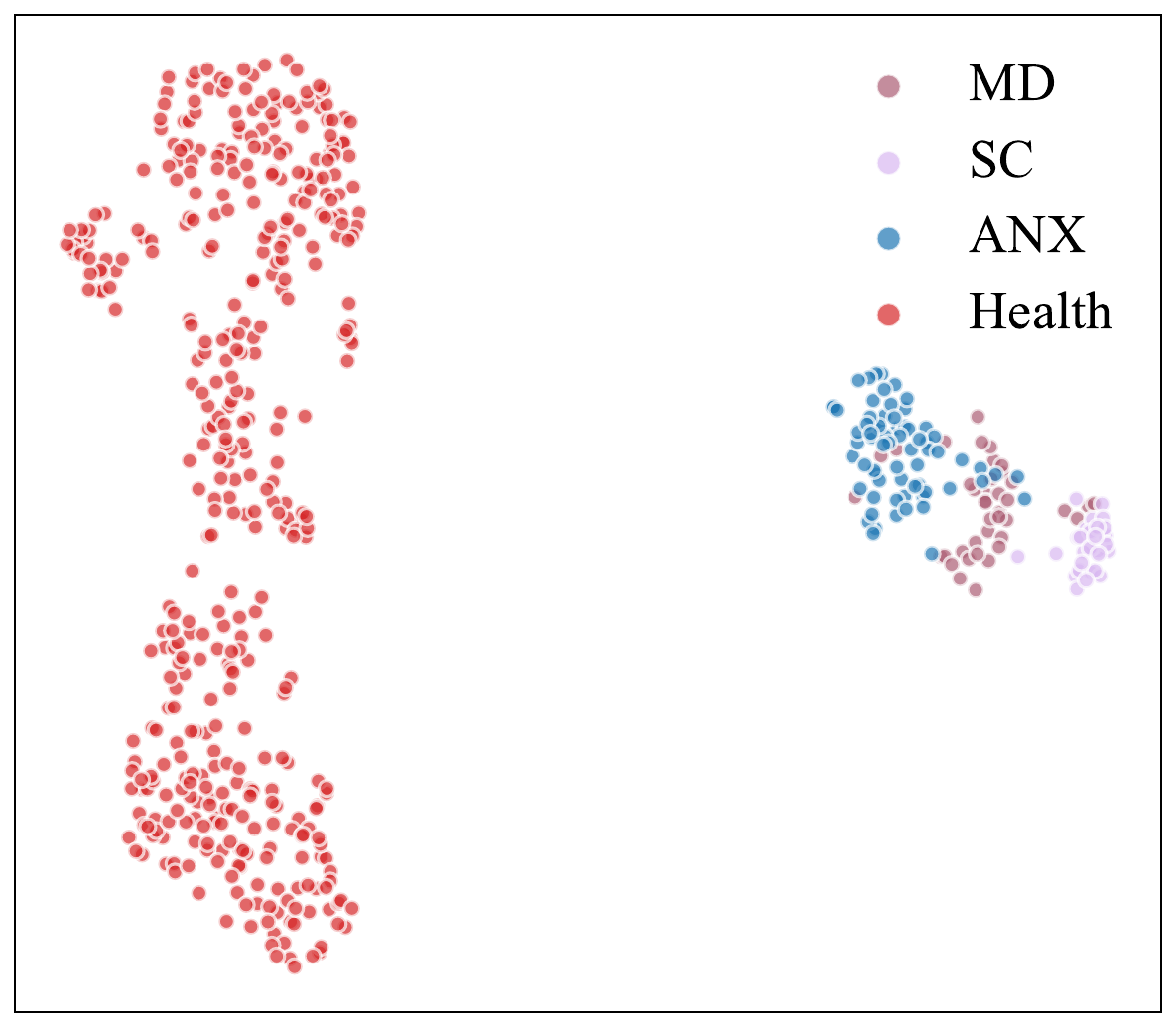}
    \hspace{-0.01\linewidth}
    \includegraphics[width=0.45\linewidth]{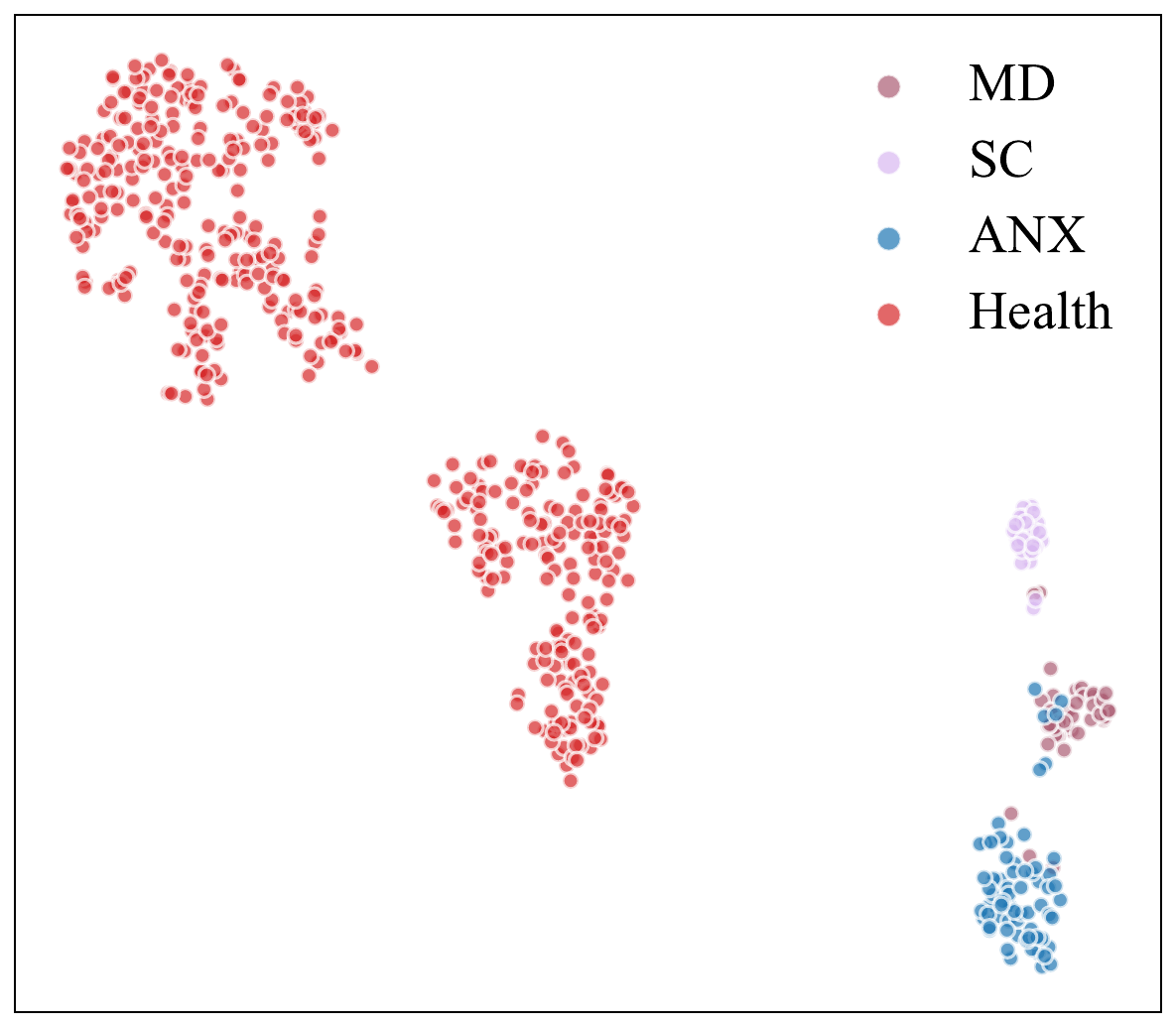}
    
    {\small (a) MD vs. SC vs. ANX vs. HC}
\end{minipage}
\hspace{0.01\textwidth}
\begin{minipage}[t]{0.45\textwidth}
    \centering
    \includegraphics[width=0.45\linewidth]{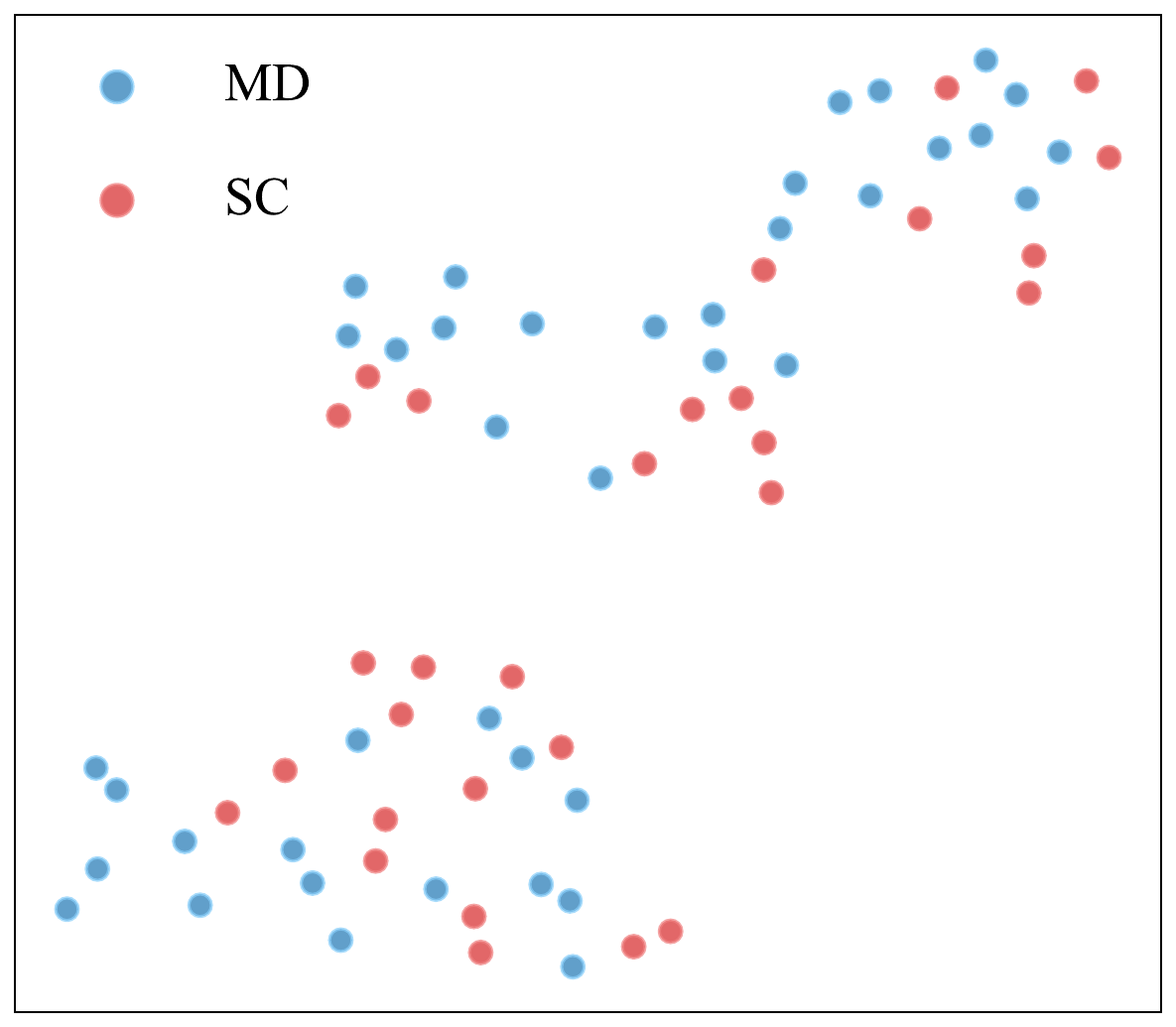}
    \hspace{-0.01\linewidth}
    \includegraphics[width=0.45\linewidth]{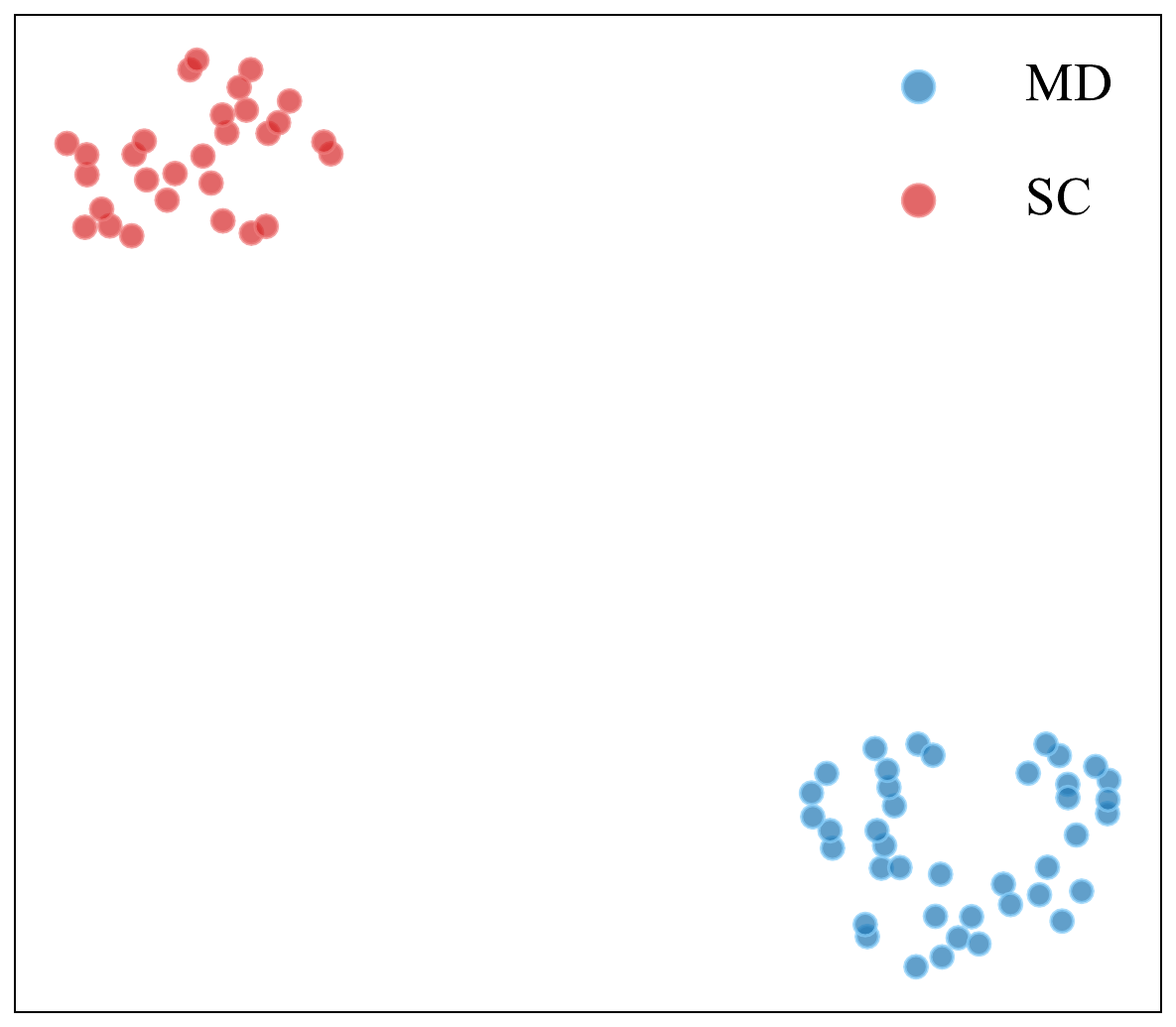}
    
    {\small (b) MD vs. SC}
\end{minipage}
\hspace{0.01\textwidth}
\begin{minipage}[t]{0.45\textwidth}
    \centering
    \includegraphics[width=0.45\linewidth]{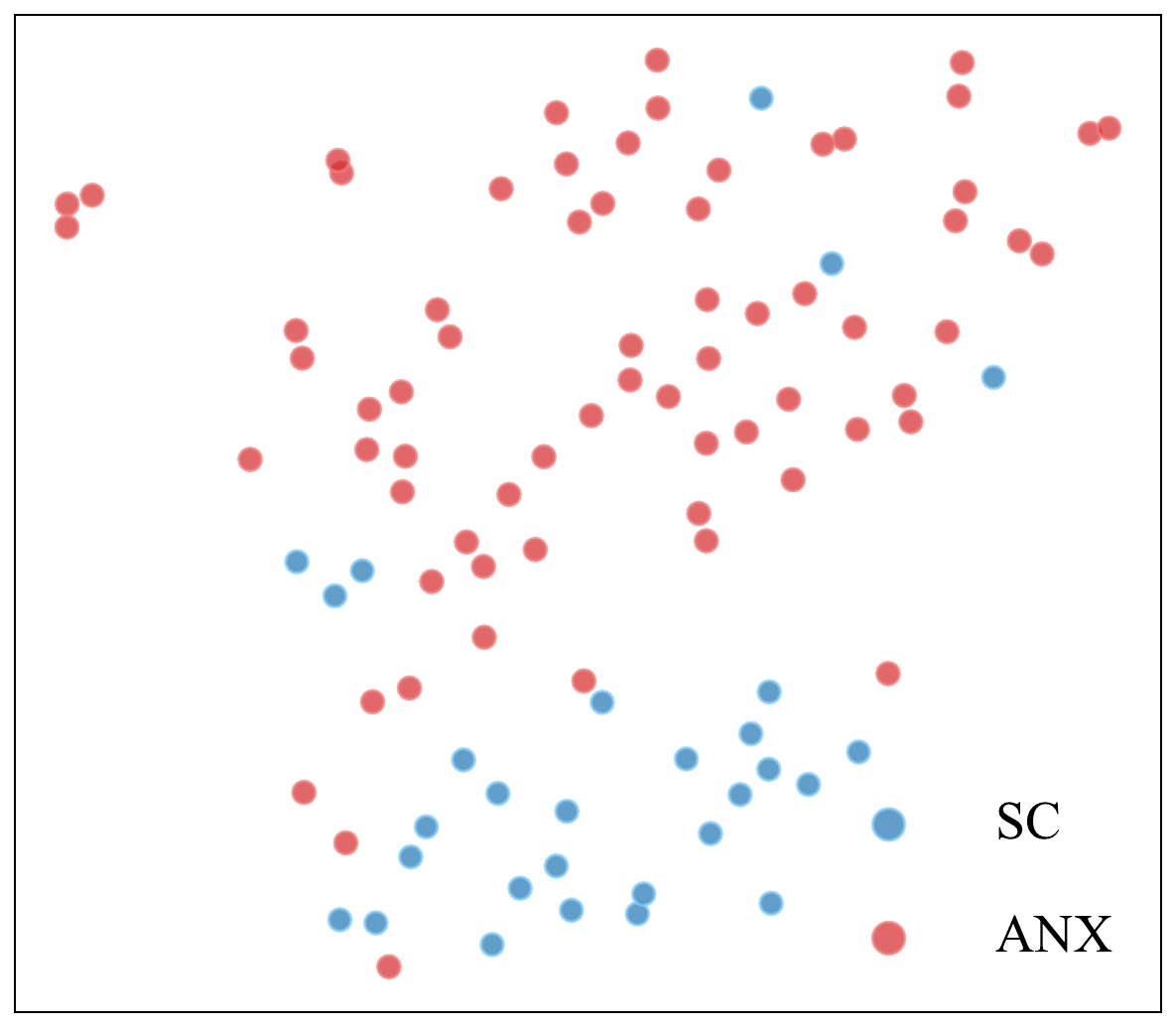}
    \hspace{-0.01\linewidth}
    \includegraphics[width=0.45\linewidth]{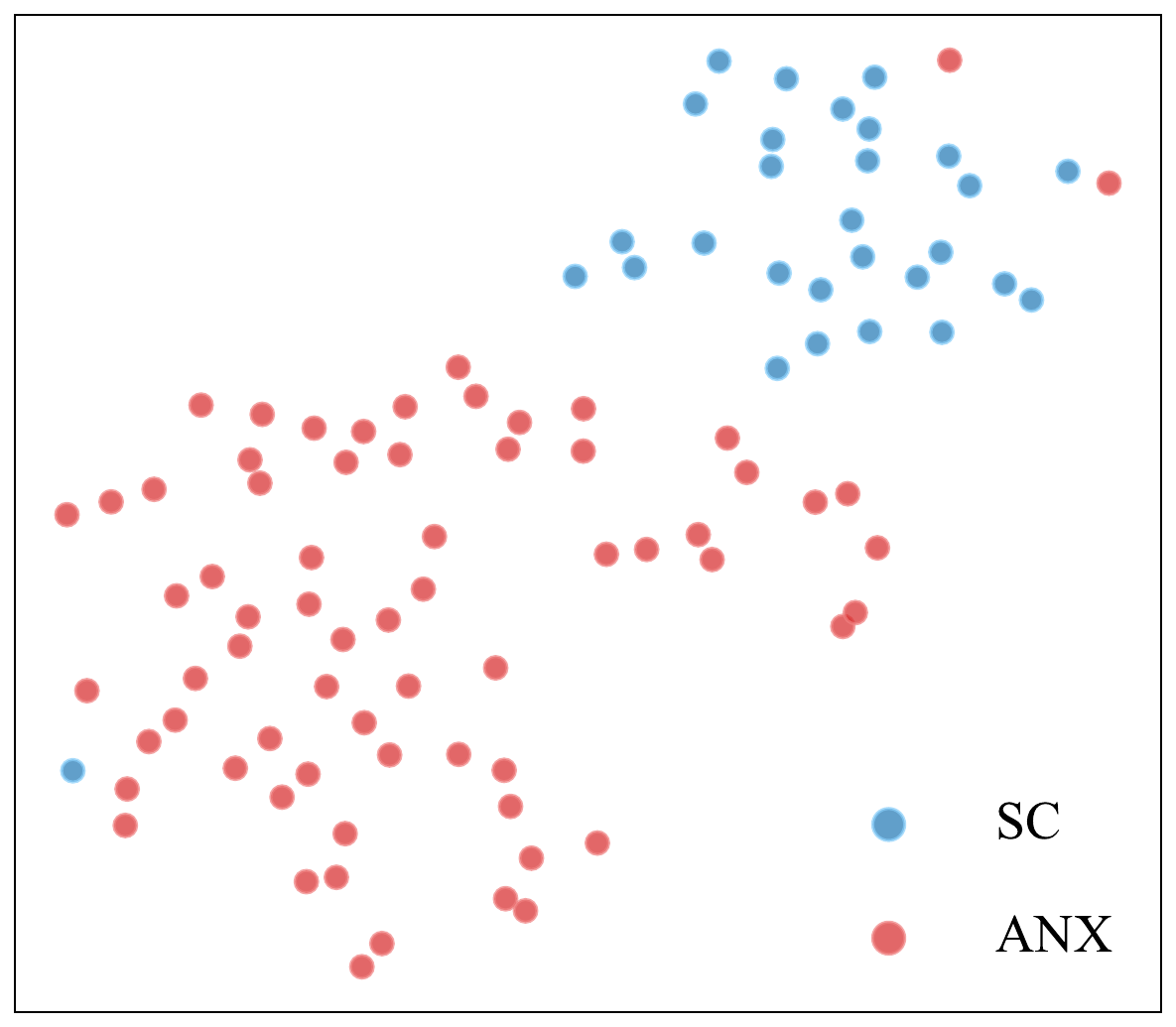}
   
    {\small (c) SC vs. ANX}
\end{minipage}

\caption{t-SNE visualization for different diagnosis tasks. Left shows the Baseline, and the right  shows Ours.}
\label{tsne}
\end{figure}

\subsection{Qualitative Interpretability}

To further examine the discriminability of the learned representations, we visualize the extracted features using t-SNE in Fig. \ref{tsne}. Compared with the baseline (introduced in \ref{fadd}), our method produces noticeably more compact intra-class clusters and clearer inter-class separation across all diagnosis settings. In the four-class setting, samples from different categories are grouped into more concentrated regions with reduced overlap, indicating that the proposed method learns more structured and discriminative representations. Similar trends can also be observed in the binary diagnosis tasks.
These results suggest that the proposed method can better capture category-specific characteristics and enhance feature compactness, thereby facilitating more reliable differential diagnosis.
The four-class setting shows clearer feature separation because it provides richer supervision than binary classification. By jointly distinguishing multiple disorders, the model learns more globally discriminative and structured representations. In contrast, binary classification only focuses on a specific pairwise boundary, leading to less organized feature distributions.

\section{Conclusions}
In this work, we take a step toward clinically realistic differential mental disorder detection. We introduce a psychology-inspired multimodal elicitation paradigm, construct a large-scale clinically verified dataset covering depression, anxiety, and schizophrenia, and propose PMLF for paradigm-aware multimodal learning across diverse elicitation tasks. Experimental results demonstrate the effectiveness of both the proposed elicitation design and modeling framework for differential mental disorder detection. One limitation of the current study is that comorbidity within the same individual is not explicitly modeled, despite its prevalence in real-world clinical settings. Future work will extend the dataset and framework toward comorbidity-aware and more comprehensive multi-disorder diagnosis scenarios.

\section*{Ethical Statement}
The data collection process has been approved by an ethics review board. Informed consent has been obtained from all participants. Personal information of the subjects is not disclosed and their extracted features of each modality are released.

\bibliographystyle{ACM-Reference-Format}
\bibliography{samples/mm}


\end{document}